\documentclass[runningheads]{llncs}
\usepackage{graphicx}
\usepackage{subcaption}
\usepackage{amsmath,amsfonts, amssymb, hyperref}
\usepackage{algorithm}
\usepackage[table]{xcolor}

\usepackage{wrapfig}
\usepackage{tikz-cd} 
\def\R{{\mathbb R}}
\def\V{{\mathcal V}}
\def\Exp{{\textrm{Exp}}}
\def\Log{{\textrm{Log}}}

\begin{document}
\title{Symmetric Algorithmic Components for Shape Analysis with Diffeomorphisms}
\titlerunning{Symmetric Shape Analysis with Diffeomorphisms}
%
\author{Nicolas Guigui\and
Shuman Jia \and
Maxime Sermesant \and
Xavier Pennec}
\authorrunning{N. Guigui et al.}
%
\institute{Universit\'e C\^ote d'Azur, Inria, Epione Project-Team, France}
\maketitle              
\begin{abstract}
In computational anatomy, the statistical analysis of temporal deformations and inter-subject variability relies on shape registration. 
However, the numerical integration and optimization required in diffeomorphic registration often lead to important numerical errors.
In many cases, it is well known that the error can be drastically reduced in the presence of a symmetry. In this work, the leading idea is to approximate the space of deformations and images with a possibly non-metric symmetric space structure using an involution, with the aim to perform parallel transport. Through basic properties of symmetries, we investigate how the implementations of a midpoint and the involution compare with the ones of the Riemannian exponential and logarithm on diffeomorphisms and propose a modification of these maps using registration errors. This leads us to identify 
transvections, the composition of two symmetries, as a mean to measure how far from symmetric the underlying structure is. We test our method on a set of 138 cardiac shapes and demonstrate improved numerical consistency in the Pole Ladder scheme.

\end{abstract}

\section{Introduction}
Computational anatomy aims at modeling the temporal evolution and cross-sectional variability of anatomical shapes. The deformations between shapes are obtained by applying non-rigid registration algorithms that seek the \textit{smallest} transformation - in a sense that will be defined precisely - of the ambient space to match two shapes. In the diffeomorphic registration setting, the deformations are modeled by diffeomorphisms, that provide invertible and folding-free transformations. 

In the Large Deformation Diffeomorphic Metric Mapping (LDDMM) framework, the space of diffeomorphisms is endowed with a right invariant metric and deformations of interest are obtained by geodesic flows from the identity transformation. They are parameterized by their initial velocity fields, which are tangent vectors at the identity deformation, defined by initial control points and dual momenta. A transformation is then computed by integration of differential equations. Registration is performed by solving an optimization problem on the initial momenta and control points with a gradient descent. These numerical schemes efficiently implement an exponential and logarithm map on a subspace of diffeomorphisms. 

However in practice, the optimization problem is relaxed to enforce smooth deformations and results in inexact matching. Therefore we can think of the exponential map as not going ``far enough''. In this work we introduce a modified exponential map that accounts for a residual error due to registration, indifferent to the choice of the regularization parameter.

The choice of the metric and regularization parameter affects the geometric structure of the space of deformations under consideration. Many convergence results depend on the curvature of this space, e.\ g.\ \cite{pennec_parallel_nodate}, and especially on its covariant derivative. This gradient being null in locally symmetric spaces, they form a very convenient setting to perform statistics on shapes. In order to assess how far from symmetric our structure is, it would therefore be valuable to develop a procedure to measure this gradient. In this paper we build on a specific parallel transport scheme.

\begin{wrapfigure}{r}{0.35\textwidth}
\centering
\vspace{-10mm}
\includegraphics[width=0.34\textwidth]{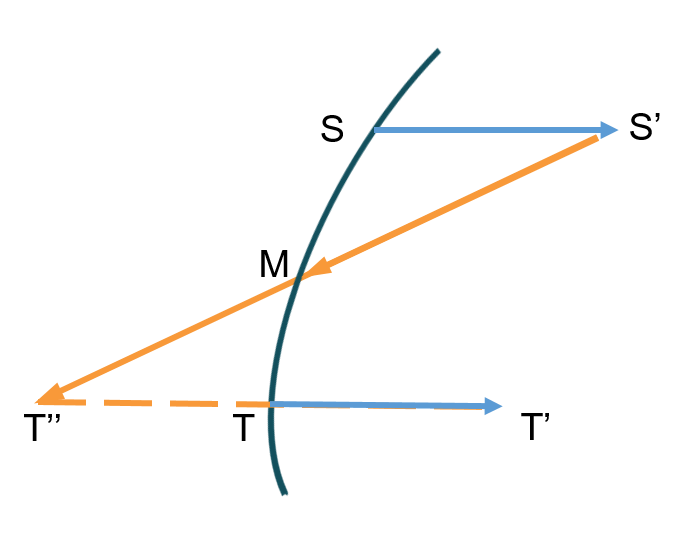}
\caption{Pole Ladder} \label{fig:pole}
\vspace{-6mm}
\end{wrapfigure}
In the statistical analysis of temporal deformations, parallel transport along geodesics is commonly used to perform inter-subject normalization, that is the vector transport of velocity fields from each subject's space to a common atlas' space. 
An approximation based on Jacobi fields was proposed in \cite{Younes_Jacobi,louis:hal-01560787}.
A numerical implementation named Pole Ladder (PL) was proposed in \cite{lorenzi:hal-00870489} and relies only on the computation of exponential and logarithm maps. Following the Shild's ladder, it consists in the construction of geodesic parallelograms. The progression between two shapes $S$ and $S'$ is transported to $T$ by:
\begin{itemize}
    \item first computing a ``midpoint'' $M$ on the geodesic between $S$ and $T$. It is seen as the diagonal of the geodesic parallelogram;
    \item then extending the geodesic from $S'$ to $M$ by the same length to obtain $T''$;
    \item similarly extending the geodesic from $T''$ to $T$ to obtain the parallel deformation of the template $T$;
\end{itemize}
See Fig.~\ref{fig:pole} for a schematic representation.

For large deformations that typically arise in inter-subject registration,
this procedure is usually iterated.
We therefore expect numerical errors to grow linearly in the number of steps, and lose crucial numerical accuracy. The accuracy of this scheme was analyzed in \cite{pennec_parallel_nodate} and shown to be a third order scheme in general, and exact in symmetric spaces where it is equivalent to a single transvection, that is, a composition of two symmetries, in our case symmetries with respect to $M$ and $T$.

In many cases, it is well known that numerical errors can be drastically reduced in the presence of a symmetry. This is for instance the case to diagonalize a symmetric matrix versus an arbitrary one. Using the Stationary Velocity Fields (SVF) framework for registration, a symmetric variant of Pole Ladder built on a Lie Group intrinsic symmetric structure was proposed in \cite{jia:hal-01860274}. This procedure is recalled in section~\ref{sec:poleLadder}. In this work, we build on this idea, but rely on LDDMM to implement a more general involution 
that accounts for the registration residual. This elementary algorithmic component constructs the symmetric shape of an original shape with respect to another one. It is presented in section~\ref{sec:residual}. In section~\ref{sec:involutivity}, we introduce the basic properties that symmetries must verify in an affine symmetric space and discuss whether these are fulfilled by our implementation. From a theoretical perspective, deviations from these properties are due to a non zero covariant derivative of the curvature tensor with the LDDMM metric. Conversely, we may interpret transvection errors as estimates of the numerical curvature gradient that encompass all the approximations due to the implementation.
The numerical experiments of section~\ref{sec:exp} show that there is an optimal regularisation parameter for which the space of deformations can best be approximated by a symmetric space. 


The paper is organised as follows: in section~\ref{sec:LDDMM}, we recall the LDDMM framework following \cite{durrleman_morphometry_2014} and in section~\ref{sec:poleLadder} the Pole Ladder procedure. We then introduce in sections~\ref{sec:residual} and~\ref{sec:involutivity} the main contributions of the paper, namely accounting for residuals, defining symmetries and their properties: centrality, involutivity and transvectivity. In section~\ref{sec:exp} we present the numerical experiments and comment on the results.

\section{Background and Method}
\label{sec:Background}

\subsection{The LDDMM Framework}
\label{sec:LDDMM}
In this work we consider shapes represented by 3D meshes. However, the methodology seamlessly applies to images. In order to define a practical finite dimensional parameterization of a subspace of diffeomorphisms $G$ acting on the ambient space $\R^d$, we consider time-varying velocity fields $v_t(x) = \sum_{k=1}^{N_c} K(x, c^{(t)}_k) \mu^{(t)}_k $ obtained by convolution of a Gaussian kernel $K(x,y) = \exp (- \frac{\Vert x-y \Vert^2}{\sigma^2})$ over $N_c$ control points $c^{(t)} = [c_k^{(t)}]_k$ and momenta $\mu^{(t)} = [\mu_k^{(t)}]_k$.

\vspace{1mm}
The set of such fields forms a pre-Hilbert space with scalar product between $v = \sum_k K(\cdot,c_k)\mu_k$ and $v' = \sum_k K(\cdot,c_k')\mu_k'$ defined by 
\begin{equation}\textstyle
    <v, v'>_H = \sum_i\sum_j K(c_i,c_j')\mu_i^T \mu_j' .
    \label{eq:prod}
\end{equation}

Diffeomorphisms are then defined as flows of velocity fields from $\phi_0 = Id$.
This amounts to integrating the ordinary differential equation (ODE) $\partial \phi_t(\cdot) = v_t[\phi_t(\cdot)]$ between $0$ and $1$.
The scalar product on velocity fields induces a right-invariant Riemannian metric on the obtained subspace of diffeomorphisms, and geodesics of this metric are parameterized by control points and momenta that satisfy the following Hamiltonian equations:
\begin{equation}
    \begin{cases}
    \Dot{c_k}^{(t)} &= \sum_j K(c_k^{(t)}, c_j^{(t)}) \mu_j^{(t)} \\
    \Dot{\mu}^{(t)} &= - \sum_j \nabla_1 K(c_k^{(t)}, c_j^{(t)}) \mu_k^{(t)^T}\mu_j^{(t)}
    \end{cases}
\end{equation}
$\phi_t$ and $v_t$ are thus uniquely determined by their initial control points $\mu$, this dependence will be explicitly written $\phi_t^{c,\mu}$ and $v_t^{c,\mu}$. In practice the interval $[0,1]$ is discretized with $n$ time steps and the ODE is solved with an iterative Euler forward or Runge-Kutta 2 method. This implements an exponential map at identity.
The registration problem between a template shape $T$ and a target mesh $S$ optimizes the following criterion over initial control points and momenta $c,\mu$:
\begin{equation}
    C(c, \mu) = \Vert S - \phi_1^{c,\mu}(T) \Vert_2^2 + \alpha^2 \cdot \| v_0^{c,\mu} \|_H^2 .
    \label{registration_crit}
\end{equation}
For simplicity, we measure the distance between shapes by the $L_2$ distance between nodes of the meshes. $\|\cdot\|_H$ is the norm defined by the scalar product of eq.~\ref{eq:prod}, which is actually the metric on $G$. The resulting $v_0$ is a logarithm at identity of $\phi_1$. In fact the metric is scaled by a factor $\alpha$, and this impacts the geometry of the underlying space as will be demonstrated. It allows to smoothly interpolate between solutions that belong to two paradigms:
\begin{itemize}
    \item 
    $\alpha \rightarrow 0$: exact matching between shapes; 
    \item 
    $\alpha \rightarrow \infty$: point distribution models (PDM), no deformation.
\end{itemize}

\subsection{Symmetric Pole Ladder for parallel transport}
\label{sec:poleLadder}
In the context of computational anatomy, the aforementioned registration framework is used to represent a subject-specific temporal deformation between shapes $S$ and $S'$, and to transport this deformation to a common atlas or template $T$ along the geodesic segment $[S,T]$. The anatomical shapes are modeled as points in a manifold $\V$ under the action of the space of diffeomorphisms $G$ described above. We suppose here that this manifold is equipped with an affine connection, which defines parallel transport and the $\Exp$ map. Locally it further defines the $\Log$ map.

Algorithm~\ref{alg:poleLadder} presents a symmetric variant of Pole Ladder \cite{lorenzi:hal-00870489} introduced in \cite{jia:hal-01860274} to approximately perform parallel transport in $\V$.
\vspace{-3mm}
\begin{algorithm}[ht]
    \caption{Mid-point symmetric Pole Ladder transport of the geodesic segment $[S, S']$ along the geodesic $[S,T]$}
    \label{alg:poleLadder}
- Compute the midpoint $M = \Exp_T(\frac{1}{2}Log_T(S))$ on the inter-subject geodesic;\\
- Compute the symmetric point $T'' = \Exp_M(-\Log_M(S'))$ of $S'$ with respect to $M$;\\
- Compute the symmetric point $T' = \Exp_T(-\Log_T(T''))$ of $T''$ with respect to $T$, and return the geodesic segment $[T, T']$.
\end{algorithm}


\vspace{-3mm}
A Taylor expansion at the midpoint $M$ of the error between the vector transported by Pole Ladder and exact parallel transport of the vector $u_S = \Log_S(S')$ is derived in \cite{pennec_parallel_nodate}. We denote by $u = \Pi_S^M u_S$ the exact transport to $M$, and  $u' = \Pi_T^M \Log_T(T')$ where $T'$ is obtained by Pole Ladder. Let also $v = \Log_M(T)$. Then
\begin{equation} 
    u' - u = \frac{1}{12}\left( (\nabla_v R)(u,v)(5 u - 2v) + (\nabla_{u} R)(u,v)(v -2 u) \right)	+ O(\|v+u\|^5).
\end{equation}
In fact this scheme is exact in an affine locally symmetric space, where $\nabla R = 0$. In this case, using the local symmetries $s_X$ at point X, we have $T' = s_T \circ s_M(S')$ meaning that Pole Ladder is equivalent to a transvection. As local symmetries are affine mappings, the following diagrams commute:
\begin{center}
    \begin{tikzcd}
         T_S \V \arrow[d, "\Pi_S^M"] \arrow[r, "(ds_M)_S"] & T_T \V \arrow[d, "\Pi_T^M"] \\
        T_M\V \arrow[r, "-Id"]  & T_M\V
    \end{tikzcd}
    \hspace{10mm}
    \begin{tikzcd}
         T_S \V \arrow[d, "\Exp_S"] \arrow[r, "(ds_M)_S"] & T_T \V \arrow[d, "\Exp_T"] \\
        \V \arrow[r, "s_M"]  & \V
    \end{tikzcd}
\end{center}
Thus, $\Pi_T^M \circ (ds_M)_S= -\Pi_S^M$ \cite[Prop 4.3]{postnikov_geometry_2001}. So with the previous notations, $(ds_M)_S u_S = -\Pi_M^T u$ and $(ds_M)_S u_S = Log_T(T") = - \Pi_M^T u'$ yielding $u' = u$ .

\subsection{Accounting for residuals to improve Centrality and Symmetry}
\label{sec:residual}


\begin{wrapfigure}{r}{0.25\textwidth}
\centering
\vspace{-10mm}
\includegraphics[width=0.2\textwidth]{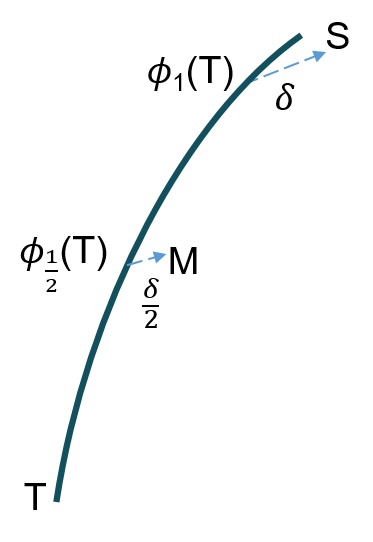}
\vspace{-4mm}
\caption{Midpoint with Residuals} \label{fig:midpoint}
\vspace{-2mm}
\end{wrapfigure}

In practice the registration is never exact due to the regularization. Thus, we propose to decompose the space of shapes into a deformation part encoding the orbit of the template, and a Euclidean space of residual displacement fields
\begin{equation}\textstyle
    S = \phi_1(T) + \delta ,
    \label{eq:decomposition}
\end{equation}
where
$\delta$ is a displacement field between corresponding points of the two meshes. Of course different possibilities exist to compute the residual, to transport it, and to apply it to different shapes, but our experiments suggest that this very simple formulation may be sufficient, so that we will not detail other approaches in this paper. 
With this decomposition, a midpoint between $T$ and $S$ is defined by
(Fig.~\ref{fig:midpoint}):
\begin{equation}
    M = \phi_{\frac{1}{2}}(T) + \frac{1}{2} \delta = \Exp_T\Big(\frac{1}{2} \Log_T(S)\Big) + \frac{1}{2} \delta .
\end{equation}
Unfortunately this formulation is not symmetric in $T$ and $S$, and registering $S$ on $T$ and shooting from $S$ results in a different midpoint in general. We will see however that using residuals decreases the distance between the midpoints obtained with the two initial points.

\begin{wrapfigure}{r}{0.30\textwidth}
\centering
\vspace{-10mm}
\includegraphics[width=0.29\textwidth]{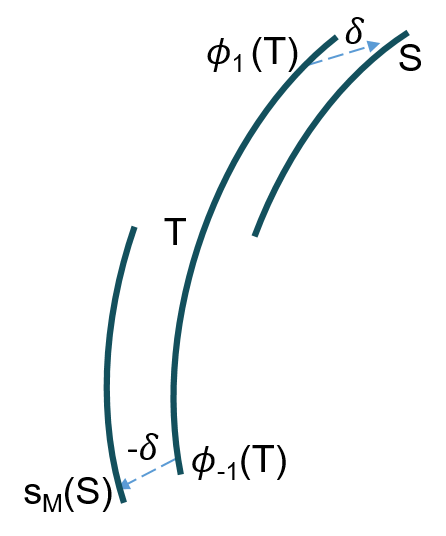}
\caption{Symmetry with Residuals} \label{fig:sym}
\vspace{-12mm}
\end{wrapfigure}


Similarly, a symmetry is defined by inverting the geodesics and the residuals (Fig.~\ref{fig:sym}):
\begin{equation}
    s_T(S) = \Exp_T \Big( - \Log_T(S)\Big) - \delta .
\end{equation}
The first desired consistency property which we refer to as \textbf{centrality} is the compatibility of the midpoint with the symmetry, namely $s_M(S) = T$. Note that this construction of a central midpoint and a local geodesic symmetry is possible in arbitrary affine connection manifolds for close enough points. However, these symmetries are affine mappings if and only if the space is locally symmetric \cite[Prop 4.2]{postnikov_geometry_2001}.

\subsection{Involutivity and Transvectivity to measure curvature gradient}
\label{sec:involutivity}
By construction our symmetry verifies $s_T(T) = T$ for all $T \in \V$, and if the $\log$ and $\exp$ maps are exact, $s_T \circ s_T = Id$. We will evaluate the exactitude of this property, called \textbf{involutivity}. Note that in a Lie Group of transformations, natural symmetries may be defined at $\phi$ by $\psi \mapsto \phi \circ \psi^{-1} \circ \phi$, which, at $\phi = Id$ is in fact the inversion. Involutivity in this case reduces to {\bf inverse consistency}: $(\psi^{-1})^{-1} = \psi \iff \psi \circ \psi^{-1} = Id$. However, in our framework, both types of errors are different because the metric exponential differs from the one defined by the canonical Cartan-Shouten connection, which is the only one compatible with the group operations.

Finally, in an affine globally symmetric space, symmetries must verify the following property that we will call \textbf{transvectivity} \cite[Prop 5.3]{postnikov_geometry_2001}:
\begin{equation}
    \forall M,S \in \V  \;\; s_M \circ s_S = s_T \circ s_M , \;\; \textrm{where} \; T=s_M(S). \label{prop:transvectivity}
\end{equation}
We want to evaluate the exactitude of this property, and use the deviation to this ideal case as a proxy to measure the gradient of the curvature of the space in the directions of interest. At this point the role of $\alpha$ becomes clearer, it allows to form a continuum of decompositions of the space of diffeomorphisms under consideration:
\begin{itemize}
    \item $\alpha \rightarrow 0, \; \delta \rightarrow 0$: the Riemannian space of deformations where the metric is not compatible with the Cartan-Shouten connection. This space is not symmetric, which shows in the transvectivity error.
    \item $\alpha \rightarrow \infty, \; v_0 \rightarrow 0$: the shapes are considered in the ambient space with Euclidean norm, this space is of course symmetric. This is the PDM framework.
\end{itemize}

We saw in section~\ref{sec:poleLadder} that Pole Ladder was doing the transvection $s_T \circ s_M$. With the same notations and using a Taylor expansion from \cite{pennec_parallel_nodate}, the deviation to parallel transport when applying the opposite transvection $s_M \circ s_S$ is:
\begin{equation}
    u'' - u = -\frac{1}{12}\left( (\nabla_v R)(u,v)(5 u + v) + (\nabla_{u} R)(u,v)(v + 2 u) \right)	+ O(\|v+u\|^5).
\end{equation}
Thus when measuring the transvectivity error $\| s_M \circ s_S(S') - s_T \circ s_M(S') \|_2$, we in fact measure $\| \Exp_T(\Pi_M^T u") - \Exp_T(\Pi_M^T u') \|_2$ where
\begin{equation}
    u' - u'' = \frac{1}{12}\left( (\nabla_v R)(u,v)(10 u - v) - 4 (\nabla_{u} R)(u,v)u \right)	+ O(\|v+u\|^5).
\end{equation}
The transvectivity error thus provides a practical way to measure the gradient of the curvature of the space even in the absence of any closed-form expression. This is noticeable in regards to the complexity of the curvature tensor itself in Mario's formula \cite{micheli_sobolev_2013} and it may lead in the future to new ways of estimating the curvature and its gradient.

\vspace{-2mm}
\section{Experiments and Application to Cardiac Shapes}
\label{sec:exp}
In this section we assess  the consistency of our numerical implementation of the symmetry compared to its theoretical properties. We compare the symmetry with residuals to standard symmetry without residuals ($\delta=0$) for different values of the parameter $\alpha$. We used Deformetrica \cite{bone_deformetrica} for all our experiments. We also compare the parallel transport obtained by Pole Ladder with both types of symmetry to the one implemented in Deformetrica \cite{louis:hal-01565478} using the fanning scheme.

\begin{figure}[!b]
\vspace{-4mm}
   \begin{subfigure}{0.32\textwidth}
        \includegraphics[height=.9\linewidth]{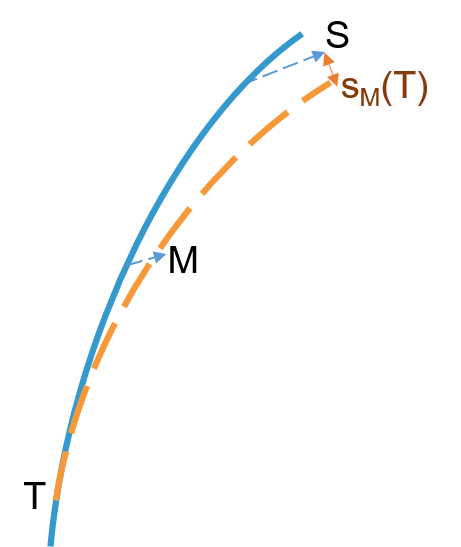}
        \caption{Centrality}\label{fig:centrality}
    \end{subfigure}
    \hspace{-15mm}
    \begin{subfigure}{0.32\textwidth}
        \includegraphics[height=.9\linewidth]{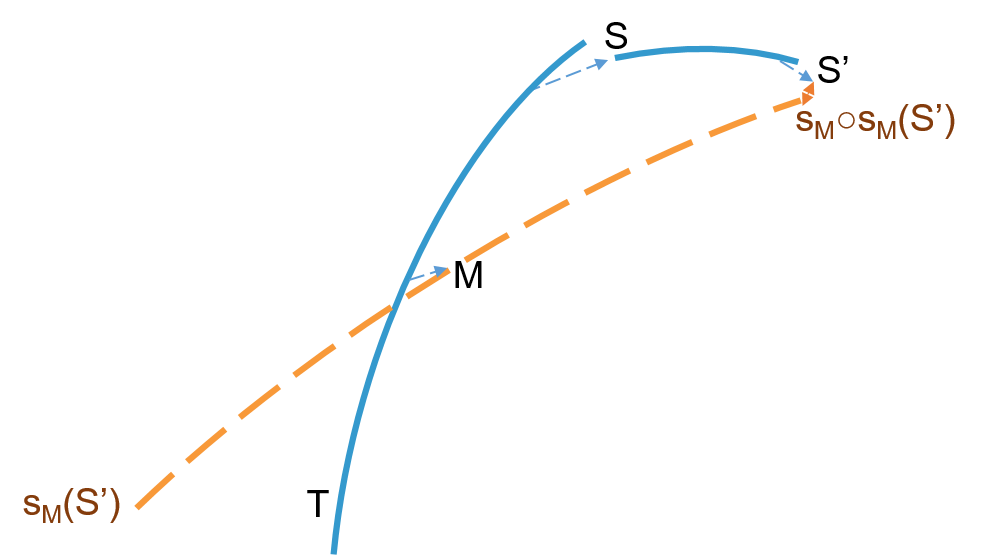}
        \caption{Involutivity}\label{fig:involutivity}
    \end{subfigure}
    \hspace{3mm}
    \begin{subfigure}{0.32\textwidth}
        \centering
        \includegraphics[height=.9\linewidth]{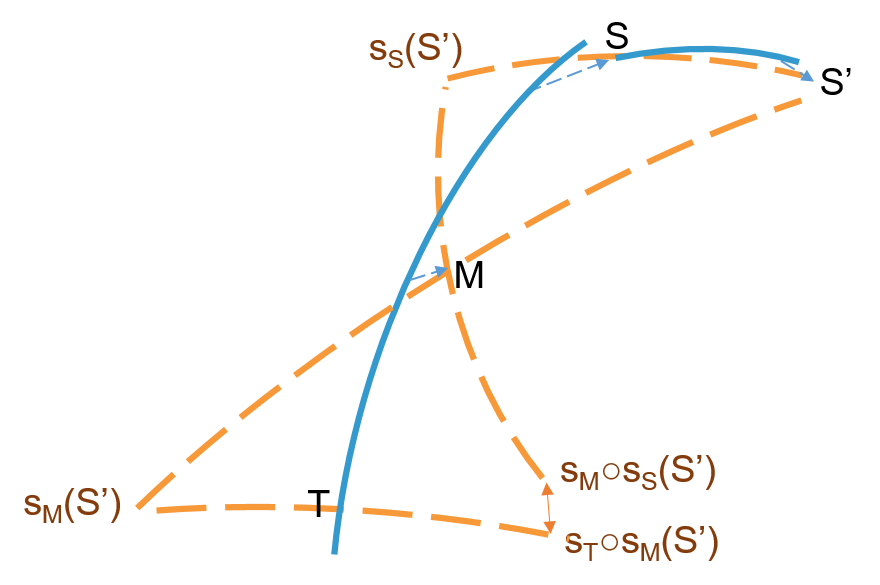}
        \caption{Transvectivity}\label{fig:transvectivity}
   \end{subfigure}
\caption{The three types of errors measured}\label{fig:errors}
\end{figure}

We use a database of cardiac shapes from 138 subjects \cite{jia:hal-01860274} for which the shapes at two time-points are available: at end-diastole ($S$) and at end-systole ($S'$). We use a population atlas as template $T$.
Four types of errors are first measured. The first is the distance between midpoints when shooting from $T$ or from $S$. The three others are, where $M$ is the midpoint obtained by shooting from $T$:
\begin{itemize}
    \item $\| s_M(T) - S \|_2$: the centrality error (Fig.~\ref{fig:centrality})
    \item $\| s_M \circ s_M(S') - S' \|_2$: the involutivity error (Fig.~\ref{fig:involutivity})
    \item $\| s_T \circ s_M(S') - s_M \circ s_S(S') \|_2$: the transvectivity error (Fig.~\ref{fig:transvectivity})
\end{itemize}

Mean results for extreme values of $\alpha$ are given in table~\ref{results}. The average registration error of $T$ on each subject's $S$, as well as the norm of the deformation and inverse consistency (by registering $S$ on $T$) for each regularisation parameter are also given for reference.
\vspace{-5mm}

\begin{table}
\vspace{-4mm}
\caption{Mean errors measured on cardiac shapes, in millimeters}\label{results}
\begin{tabular}{ |c||c|c|c|c|c|c|}  
 \hline & 
 \multicolumn{2}{|c|}{$\alpha^2=0.01$} &
 \multicolumn{2}{|c|}{$\alpha^2=1$} & 
 \multicolumn{2}{|c|}{$\alpha^2=1089$}\\
 \hline
 Error Type & Residual & No Residual & Residual & No Residual & Residual & No Residual\\
 \hline
 Centrality & 0.36& 0.43& 0.10& 0.50& $< 0.01$& 5.76\\
 Involutivity & 1.42& 1.55& 0.33& 0.80& $< 0.01$& 9.39\\
 Transvectivity & 1.98& 2.16& 0.58& 0.62& $< 0.01$& 0.16 \\
 \hline
 Reg.\ Error &
 \multicolumn{2}{|c|}{0.23} &
 \multicolumn{2}{|c|}{0.41} & 
 \multicolumn{2}{|c|}{5.61}\\ \hline
 Reg.\ Norm &
 \multicolumn{2}{|c|}{42} &
 \multicolumn{2}{|c|}{30} & 
 \multicolumn{2}{|c|}{1}\\ \hline
 Inverse Cons.\ &
 \multicolumn{2}{|c|}{0.13} &
 \multicolumn{2}{|c|}{0.14} & 
 \multicolumn{2}{|c|}{0.10}\\
 \hline
\end{tabular}
\vspace{-4mm}
\end{table}

\begin{figure}[ht]
\vspace{-5mm}
    \begin{subfigure}{0.49\textwidth}
        \includegraphics[height=0.82\textwidth]{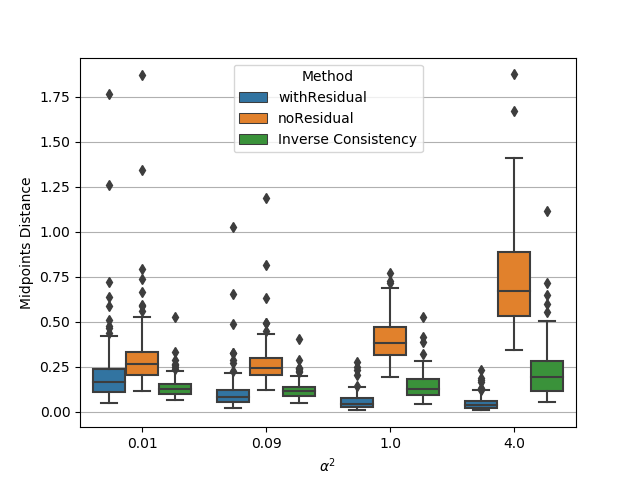}
        \caption{Distance between Midpoints}
        \label{fig:bp-midpoints}
    \end{subfigure}
    \begin{subfigure}{0.49\textwidth}
        \includegraphics[height=0.82\textwidth]{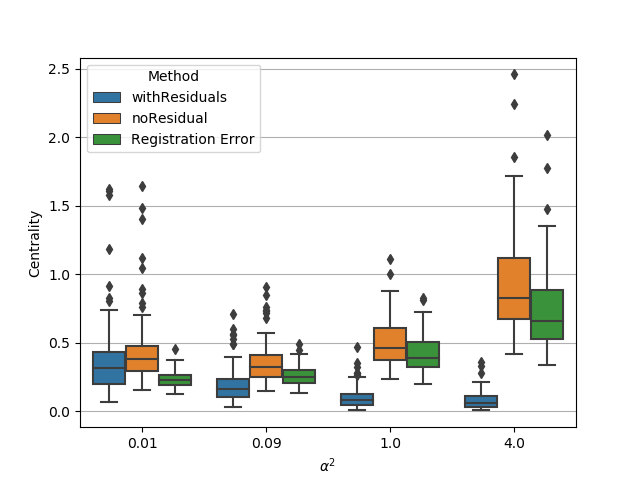}
        \caption{Centrality}
        \label{fig:bp-centr}
    \end{subfigure}
    \vspace{-3mm}
    \begin{subfigure}{0.49\textwidth}
      \includegraphics[height=0.82\textwidth]{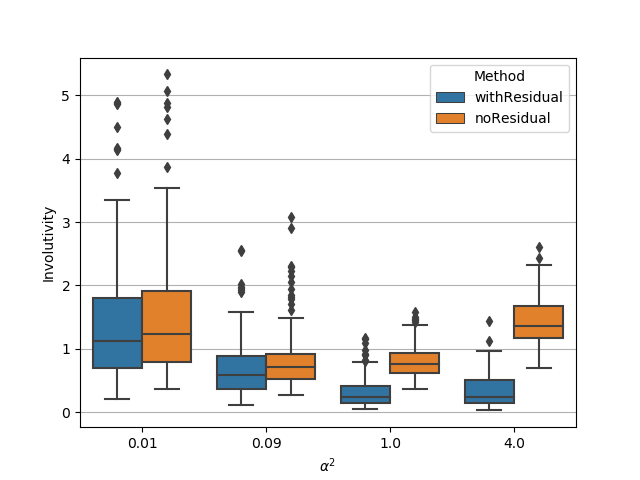}
      \caption{Involutivity}
      \label{fig:bp-inv}
    \end{subfigure}
    \begin{subfigure}{0.49\textwidth}
        \includegraphics[height=0.82\textwidth]{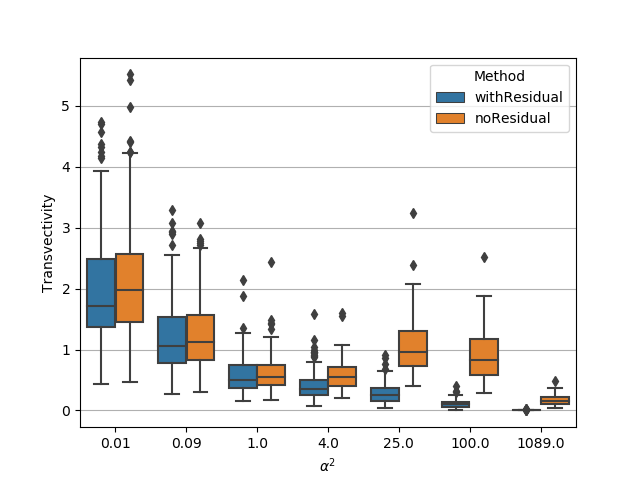}
        \caption{Transvectivity}
        \label{fig:bp-trans}
    \end{subfigure}
    \caption{The four errors for different values of $\alpha^2$, with and without residuals in the computation of the midpoint and the symmetries.}
    \vspace{-3mm}
    \label{fig:bp-errors}
\end{figure}

These results illustrate the two contributions of this paper. Firstly, using residuals in the symmetry considerably improves the numerical accuracy of the computation of a midpoint. As we can see on Fig.~\ref{fig:bp-midpoints}, the distance between midpoints computed by shooting from $T$ or from $S$ is reduced when using the residuals. This error compares well with the inverse consistency error in general and is even significantly lower for $\alpha \geq 1$.

Moreover, this increase in numerical accuracy is also visible on the centrality and involutivity errors. Indeed, for centrality (Fig.~\ref{fig:bp-centr}), the error when using residuals is consistently smaller than without residuals, and this gain becomes larger as $\alpha$ grows. It is also remarkable that this error is significantly lower than the registration error for $\alpha \geq 1$. The same behavior is observed for the involutivity on Fig.~\ref{fig:bp-inv}. This means that for reasonable values of $\alpha$ we obtain reliable implementations of the midpoint and symmetry.

Secondly, the LDDMM space endowed with the right invariant metric is not symmetric, and the transvectivity error reflects the covariant derivative of the curvature. Fig.~\ref{fig:bp-trans} gives further details: pushing registration with a small $\alpha$ generates larger errors, and there is an optimal $\alpha \in [1;2]$ for which the space is closest to being symmetric. Using residuals, this error decreases as the space flattens to a Euclidean space and deformations tend to the identity.


\begin{figure}[ht]
\vspace{-5mm}
  \begin{subfigure}{0.55\textwidth}
      \includegraphics[width=0.90\textwidth]{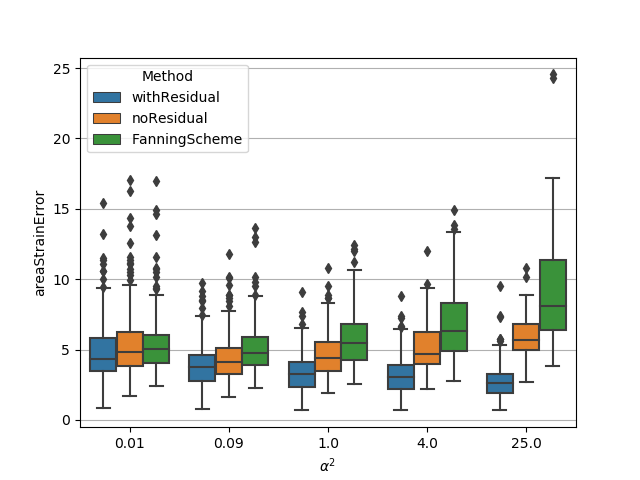}
      \caption{Local Area Change errors for different values of $\alpha$ with and without residuals in Pole Ladder, and with Deformetrica's fanning scheme.}
      \label{fig:area}
  \end{subfigure}
  \begin{subfigure}{0.45\textwidth}
      \includegraphics[width=0.85\textwidth]{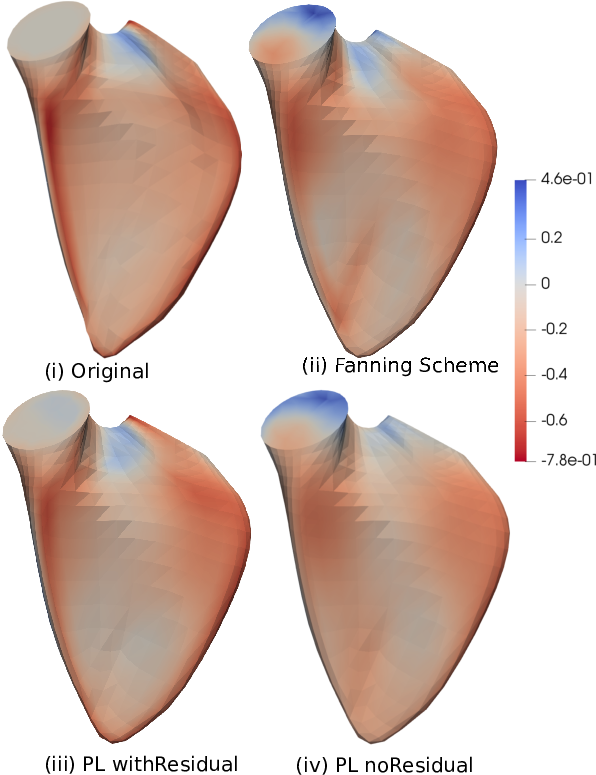}
      \caption{Area Strain maps}
      \label{fig:areamaps}
  \end{subfigure}
  \caption{b) Area strain maps for one patient computed between S and S' and represented on S (top left), and computed between T and T' and represented on T, where T' is obtained with the different methods and $\alpha=1$.}
  \label{fig:areachanges}
  \vspace{-5mm}
\end{figure}

Finally, in order to evaluate the result of symmetric Pole Ladder and compare it to the fanning scheme, we compute the local area strain (LAS) between end-diastole and end-systole at every landmark of the mesh. For two corresponding points $m_i,m_i'$ that belong to $k_i$ triangular cells, we compute the mean of the difference of area of each of these cells between $S$ and $S'$. 
\begin{equation}
    LAS_i^{S-S'}= \frac{1}{k_i} \sum_{j=1}^{k_i} \frac{(a_j - a_j')}{a_j}
\end{equation}
This feature is commonly used by clinicians to characterise the cardiac motion \cite{kleijn_three-dimensional_2011}. Here we use it to test the isometric property of the parallel transport scheme: we measure the area strain between the original subject's meshes $S$ and $S'$, and compare it to the one measured between the atlas $T$ and the deformed atlas $T'$ obtained by the parallel transport algorithm. We report in Fig.~\ref{fig:area} the square root of the sum of squared differences over all landmarks:
\begin{equation}
    ASE^2 = \sum_{i} (LAS_i^{S-S'} - LAS_i^{T-T'})^2
\end{equation}

As $\alpha$ grows, both the temporal and subject-to-atlas deformations decrease, thus generating less area changes, which explains the growing errors. Furthermore, we can see on Fig.~\ref{fig:areamaps} that the area strain is dominated by a bending artifact of the valve sections at the borders. Using residuals reduces this effect in this example. Although more suitable methods may exist to directly map scalar functions from one shape to another, these results emphasize the contribution of this paper: using residuals improves the symmetry and parallel transport with Pole Ladder.

\vspace{-5mm}
\section{Conclusion}
\vspace{-2mm}
We introduced residuals from registration errors to compute midpoints and symmetries between shapes. This results in improved numerical consistency for the centrality and involutivity properties. Furthermore the transvectivity error reflects the curvature of the underlying Riemannian space of deformations, and allows to estimate how far from symmetric this space is, depending on the regularisation parameter. Performing the same experiments in the framework of SVF with a similar implementation would yield very interesting comparison as the space is naturally symmetric, and accounting for residuals would thus provide a more consistent method for parallel transport. This gain could be even more interesting when considering registration between images. Indeed images are not in the template's orbit in practice and residuals would encode intensity bias, which is a key source of error in image registration.

\vspace{-1mm}
\paragraph{Acknowledgements:} This project has received funding from the European Research Council (ERC) under the European Union’s Horizon 2020 research and innovation program (grant agreement G-Statistics No 786854).
%
%
%
\bibliographystyle{splncs04}
\vspace{-4mm}
\bibliography{biblio.bib}

\end{document}